\documentstyle[din_a4,12pt]{article}
\begin{document}
\setlength{\baselineskip}{0.65cm}
\setlength{\parskip}{0.5cm}
\begin{titlepage}
\begin{flushright}
MPI-PhT/94-22 \\ LMU--3/94 \\ April 1994
\end{flushright}
\vspace{0.5cm}
\begin{center}
\Large
{\bf Light Gluinos and the Parton Structure of the Nucleon$\, ^{\ast}
$} \\
\large
\vspace{2.0cm}
R. R\"{u}ckl$\, ^{a,b}$ and A. Vogt$\, ^{a}$ \\
\vspace{1.5cm}
\normalsize
{$ \mbox{}^{a} $
 Sektion Physik, Universit\"at M\"unchen, Theresienstra{\ss}e 37} \\
{ D--80333 Munich, Germany} \\
\vspace{0.5cm}
{$ \mbox{}^{b} $
 Max-Planck-Institut f\"ur Physik, Werner-Heisenberg-Institut,
F\"ohringer Ring 6} \\
{ D--80805 Munich, Germany} \\
\vspace{2.0cm}
\large
{\bf Abstract} \\
\end{center}
\vspace{-0.3cm}
\normalsize
We study the effects of light gluinos with $ m_{\tilde{g}} \stackrel{<}
{{\scriptstyle \sim}} 1 \mbox{ GeV} $ on the nucleon parton
densities and the running of $ \alpha _{S} $. It is shown that from
the available high-statistics DIS data no lower bound on the gluino
mass can be derived. Also in the new kinematical region accessible at
HERA the influence of such light gluinos on structure functions is
found to be very small and difficult to detect. For use in more direct
searches involving final state signatures we present a radiative
estimate of the gluino distribution in the nucleon.
\vfill
\noindent
$^{\ast} $ Work supported by the German Federal Ministry for Research
and Technology under contract No.\ 05 6MU93P
\end{titlepage}
\section{Introduction}
The values of the strong coupling constant $ \alpha _{S}(Q) $ as
determined from fixed-target deep inelastic lepton-nucleon scattering
(DIS) and from hadronic decays of the $Z$-boson at LEP slightly
deviate from the running predicted by QCD \cite{Beth1,Beth2}. This
observation has kindled renewed speculations about the possible
existence of a light gluino \cite{FF} with a mass of at most a few GeV
\cite{AEN,Clav,JK,ENR}. The presence of such a light gluino would
slow down the decrease of $ \alpha _{S}(Q) $ between the average
scale $ Q \simeq 5 \mbox{ GeV} $ of the DIS measurements and the
$Z$-mass and thus improve the overall agreement of experiment with
theory \cite{JK,ENR}.
The latest $ \alpha _{S} $-determinations have reduced the original
discrepancy mainly because of a slightly smaller hadronic width of the
$Z$ \cite{Beth2}. Within errors, the observed running is now consistent
with standard QCD as well as with the hypothesis of a light majorana
gluino. Since also other measurement do not convincingly rule out light
gluinos \cite{PDG}, this chapter is not yet closed.

If a light gluino exists, it does not only modify the running of
$ \alpha _{S}(Q) $, but also the evolution of parton densities and DIS
structure functions \cite{CER,AKL}.  It is therefore interesting to
investigate the question whether or not one can derive lower bounds on
the gluino mass from present or future structure function measurements
alone. Such a bound could not be undermined by theoretical
uncertainties and would thus put an unavoidable constraint.
Following ref.\ \cite{UA1} it is reasonable to concentrate on the mass
windows \ $ m_{\tilde{g}} \stackrel{<}{{\scriptstyle \sim}} 1 \mbox{
GeV} $ and $ 3 \mbox{ GeV} \stackrel{<}{{\scriptstyle \sim}}
m_{\tilde{g}} \stackrel{<}{{\scriptstyle \sim}} 5 \mbox{ GeV} $. Gluinos
in the upper mass range do not significantly change the successful
description of fixed-target DIS, direct photon and Drell-Yan lepton
pair production obtained by global fits \cite{MRS,CTEQ} in the
framework of the QCD improved parton model.  Furthermore, also
extrapolations of DIS structure functions to the new kinematical regime
accessible at HERA show only little sensitivity to a gluino in this
mass range \cite{RS}.
However, it is a priori not obvious whether these conclusions also
apply to the lower mass window $ m_{\tilde{g}} < 1 \mbox{ GeV} $.
Such very light gluinos would carry as much as about 10\% of the
nucleon momentum already at scales probed in fixed-target DIS.
Thus the task remains to clarify whether existing DIS data (including
first HERA results at small $x$ \cite{ZEUS,H1}) rule out the
existence of light gluinos below 1 GeV. If this is not the case, the
question arises whether one can expect observable effects later at
HERA when more statistics will be accumulated in particular at large
$x$ and large $ Q^{2} $.

Some of these issues have been investigated in a very recent analysis
\cite{BB} following the procedure of the CTEQ collaboration \cite{CTEQ}.
It is concluded that the fixed-target data do not discriminate between
the presence and absence of light gluinos. On the other hand, the
scaling violations at $ x \stackrel{<}{{\scriptstyle \sim}} 10^{-2}
$ are found to be distinctly different in these two cases, whence
small-$x$ data from HERA should soon provide decisive tests. This
finding, however, contradicts the results of ref.\ \cite{RS} for
the common mass $ m_{\tilde{g}} = 5 \mbox{ GeV} $.

In the present paper, we study the questions raised above within the
framework of radiatively generating the parton densities from low-scale
valence-like input distributions \cite{GRV2}. This approach is
especially well suited for the incorporation of a new particle species
with a mass below the input scale $ Q_{0} \simeq 2 \mbox{ GeV} $ of
conventionally fitted parton distributions \cite{MRS,CTEQ}, since it
avoids the introduction of an additional unknown input function. For
the gluino, this initial distribution cannot be well determined by a
global fit, while our procedure leads to a rather definite prediction.
This radiative estimate of the gluino density in the proton is useful
for the study of more direct searches using final state signatures
\cite{JK,ENR,BB}.
As far as DIS structure functions are concerned, we find no significant
effects from gluinos with $ m_{\tilde{g}} \stackrel{<}{{\scriptstyle
\sim}} 1 \mbox{ GeV} $, neither in the kinematical regime explored so
far, nor in the new region accessible to HERA. Thus our analysis does
not confirm the low-$x$ results of ref.\ \cite{BB} mentioned above.

The paper is organized as follows:
In section 2, we recall the formalism for including a light
gluino in the $ Q^{2} $-evolution of hadronic parton densities
\cite{CER,AKL} and derive the analytical Mellin-$n$ space solution
of the modified evolution equations. In section 3, we then explain
our phenomenological procedure and present fits to a suitably chosen
subset of the available high-statistics DIS measurements. The emerging
parton distributions and their implications are discussed in section 4.
Finally, in section 5,  we summarize our results.
\section{Theoretical Framework}
In this section we formulate the evolution of parton densities in
next-to-leading order (NLO) in the presence of a light majorana gluino.
The corresponding standard QCD results can be recovered by simply
dropping the gluino-gluon and gluino-quark splitting functions together
with the gluino effects in the usual splitting functions.
All expressions will be given in $n$-moment space defined by
the Mellin transformation
  \begin{equation}
    a^{n} = \int_{0}^{1}\! dx\,x^{n-1} a(x) \:\: .
  \end{equation}
In this representation, the convolutions in evolution equations and
structure functions reduce to ordinary products. As a consequence, the
equations can be solved analytically. The inverse Mellin transformation
back to Bjorken-$x$ space has then to be performed numerically.

The strong coupling constant in NLO can be written in the form
  \begin{equation}
    \frac{\alpha_{S}(Q^{2})}{4\pi } = \frac{1}{\beta_{0} \ln
      (Q^{2}/\Lambda^{2})} - \frac{\beta_{1}}
      {\beta_{0}^{3}} \frac{\ln \ln (Q^{2}/\Lambda^{2})}
      {[\ln (Q^{2}/\Lambda^{2})]^{2}} \:\: .
  \end{equation}
The (scheme-independent) coefficients of the $\beta $-function are
given by
  \begin{equation}
    \beta_{0} = 11 - (2/3)f \: , \:\:\: \beta_{1} = 102 - (38/3) f
  \end{equation}
in the standard QCD case, and by
  \begin{equation}
    \beta_{0} = 9 - (2/3)f \: , \:\:\:  \beta_{1} = 54 - (38/3) f
  \end{equation}
in the presence of a majorana gluino. Here $f$ denotes the number of
relevant quark flavours. The QCD scale parameter $ \Lambda $ depends on
this number as well as on the renormalization scheme. For the present
analysis we employ, as usual, the $\overline{\mbox{MS}} $ scheme.
As can be seen from eqs.\ (3) and (4), a light  gluino slows down the
running of $ \alpha_{S}(Q^{2}) $.

Beyond the leading logarithmic approximation  parton densities
and their evolution depend on the factorization scheme, i.e.\ on the
definition of the coefficient functions $ C_{i} $ entering
the moments of the familiar structure functions $F_{1}$ and $F_{2}$:
  \begin{equation}
    (\delta_{i} F_{i})^{n}(Q^{2}) = \sum_{k=1}^{f} e_{k}^{2} \bigg\{
      \bigg( 1+ \frac{\alpha_{S}(Q^{2})}{2\pi}C_{i,q}^{n} \bigg)
      [ q_{k}^{n}(Q^{2}) + \bar{q}_{k}^{n}(Q^{2}) ]
      + \frac{\alpha_{S}(Q^{2})}{2\pi} \frac{1}{f} C_{i,G}^{n}
      G^{n}(Q^{2}) \bigg\} \:\: ,
  \end{equation}
where $ \delta_{1} = 2 $, $ \delta_{2} = 1/x $. $ (\bar{q}_{k}) \:
{q}_{k} $ and $G$ denote the (anti-)quark and gluon distributions,
respectively, and $e_{k}$ is the electric charge in units of the
electron charge. In contrast to ref.\ \cite{AKL}, we adopt the
$\overline{\mbox{MS}} $ scheme \cite{BaBu}. This allows for a direct
comparison of the resulting parton densities with distributions from
conventional QCD analyses. Since the gluino coupling to electroweak
gauge bosons is $ {\cal O} (\alpha_{S}^{2}) $, the NLO coefficient
functions $ C_{i}^{n} $ are the same with and without a gluino. The
explicit expressions in  $\overline{\mbox{MS}} $ can be found in
ref.\ \cite{FKL}.

In the $\overline{\mbox{MS}} $ scheme the massive quark flavours and
the gluino can be included in the evolution as follows \cite{CT}:
Below the threshold $ Q^{2} = m_{h}^{2} $ the heavy degree of freedom
$h$ is neglected in the evolution of parton densities as well as in
the coefficients (3) and (4) of the $\beta $-function. At
$ Q^{2} > m_{h}^{2} $ it participates in the evolution and the running
of $ \alpha_{S}(Q^{2}) $ like a massless flavour but with the boundary
condition
  \begin{equation}
    h^{n}(m_{h}^{2}) = 0 \:\: .
  \end{equation}
This guarantees the continuity of the parton densities at threshold.
The $\overline{\mbox{MS}} $ scale parameter $ \Lambda^{(f)} $ is
changed at $ Q^{2} = m_{h}^{2} $ by the amount necessary to ensure
continuity of  $ \alpha_{S}(Q^{2}) $.

In the presence of a gluino the evolution equations read
\begin{eqnarray}
  \frac{d}{d\ln Q^{2}} v^{n}_{\eta ,i}(Q^{2})
    &=& \frac{\alpha_{S}(Q^{2})}{2\pi } \left( P_{qq}^{(0)n} +
	\frac{\alpha_{S}(Q^{2})} {2\pi} P^{(1)n}_{\eta}
        \right) v_{\eta ,i}^{n}(Q^{2})                   \nonumber \\
  \frac{d}{d\ln Q^{2}}\vec{q}^{\,n}(Q^{2})
    &=& \frac{\alpha_{S}(Q^{2})}{2\pi } \left( \hat{P}^{(0)n} +
	\frac{\alpha_{S}(Q^{2})} {2\pi} \hat{P}^{(1)n}
        \right) \vec{q}^{\,n}(Q^{2}) \:\: .
\end{eqnarray}
We have simplified the equations by introducing the usual
non-singlet combinations of quark flavours ($ i = 1,\ldots ,f $),
  \begin{equation}
     v_{-,i}^{n} = q_{i}^{n} - \bar{q}_{i}^{n}  \:\: , \:\:\:
     v_{+,i^{2}-1}^{n} = \sum_{k=1}^{i} (q_{k}^{n} + \bar{q}_{k}^{n})
                  -i (q_{i}^{n} + \bar{q}_{i}^{n}) \:\: ,
  \end{equation}
together with the vector of singlet moments
  \begin{equation}
    \vec{q}^{\,n} = \left( \begin{array}{c} \Sigma^{n} \\ G^{n} \\
		    \tilde{g}^{n} \end{array} \right) \:\: , \:\:\:
    \Sigma^{n} = \sum_{i=1}^{f} (q_{i}^{n} + \bar{q}_{i}^{n}) \:\: ,
  \end{equation}
$f$ being again the number of active quark flavours.
The moments of the splitting functions are generically denoted by
$ P^{(0)n} $ and $ P^{(1)n} $ in first and second order, respectively.
Note that no additional non-singlet combination involving the gluino
distribution $ \tilde{g}^{n}(Q^{2}) $ occurs since we only consider
majorana gluinos here.

The splitting functions in eq.\ (7) can be expressed in terms of the
standard QCD quark and gluon splitting functions by a change of colour
factors. In lowest order, $ P_{qq}^{(0)n} $ is unchanged, while the
modified singlet matrix $ \hat{P}^{(0)n} $ is given by \cite{CER}
  \begin{equation}
   \hat{P}^{(0)n} = \left( \begin{array}{ccc}
   P^{(0)n}_{qq} & 2fP^{(0)n}_{qG} &      0                   \\[0.1cm]
   P^{(0)n}_{Gq} & P^{(0)n}_{GG}-1 & \frac{9}{4}P^{(0)n}_{Gq} \\[0.1cm]
        0        & 6P^{(0)n}_{qG}  & \frac{9}{4}P^{(0)n}_{qq}
                           \end{array} \right) \:\: .
  \end{equation}
Note the absence of quark-gluino splitting in LO and the modification
of the gluon-gluon splitting function by the gluino contribution to
the gluon self-energy graph. The main effect of the gluino on the
quarks and gluons is a reduction of the gluon density. The second order
quantities $ P_{\eta}^{(1)n} $ and $ \hat{P}^{(1)n} $ follow from those
without a gluino by changing the group factors as described in
table 1 of ref.\ \cite{AKL}\footnote{There are two misprints in table
1 of ref.\ \cite{AKL}: the $ C_{F}N_{F} $ entries of GL and NS$_{L}$
should both read $ C_{\lambda} (N_{F}+N_{\lambda}) $.}.
The explicit expressions for integer values of $n$ can be inferred
from the appendix of ref.\ \cite{FKL} using $ P^{(0)n} = - \gamma
^{(0)n} /4 $ and $ P^{(1)n} = - \gamma ^{(1)n} /8 $.
The analytic continuation to complex $n$, necessary for the Mellin
inversion of eq.\ (1), is explained in ref.\ \cite{GRV1}.

The solution of the non-singlet part of eq.\ (7) is straightforward.
One finds
  \begin{equation}
  v_{\eta ,i}^{n}(Q^{2}) = \left\{ 1 - \frac{\alpha_{S}(Q^{2}) -
    \alpha_{S}(Q_{0}^{2})}{\pi \beta_{0}}  \bigg( P_{\eta}^{(1)n}
    - \frac{\beta_{1}}{2\beta_{0}} P_{qq}^{(0)n} \bigg) \right\}
    \left( \frac{\alpha_{S}(Q^{2})}{\alpha_{S}(Q^{2}_{0})} \right)
    ^{-(2/\beta_{0}) P_{qq}^{(0)n}} v_{\eta ,i}^{n}(Q^{2}_{0}) \:\: .
  \end{equation}
In the singlet case, the situation is more complicated due to the
non-commutativity of the matrices $ \hat{P}^{(0)} $ and $ \hat{P}^{(1)}
$. An implicit solution is given by
\begin{eqnarray}
  \vec{q}^{\,n}(Q^{2}) &=&
  \left\{ \left[ \alpha_{S}(Q^{2})/\alpha_{S}(Q^{2}_{0}) \right]
  ^{-(2/ \beta_{0}) \hat{P}^{(0)n}} + \frac{\alpha_{S}(Q^{2})}{2\pi}
  \hat{U}^{n} \left[ \alpha_{S}(Q^{2})/\alpha_{S}(Q^{2}_{0}) \right]
  ^{-(2/ \beta_{0}) \hat{P}^{(0)n}} \right.              \nonumber \\
  & & \mbox{} \left. - \frac{\alpha_{S}(Q_{0}^{2})}{2\pi}
  \left[  \alpha_{S}(Q^{2})/\alpha_{S}(Q^{2}_{0}) \right]
  ^{-(2/ \beta_{0}) \hat{P}^{(0)n}} \hat{U}^{n} \right\}
  \vec{q}^{\,n}(Q_{0}^{2}) \:\: ,
\end{eqnarray}
where the subleading evolution matrix $ \hat{U} $ is fixed by the
commutation relation
  \begin{equation}
     [ \hat{U}^{n}, \hat{P}^{(0)n} ] = \frac{\beta_{0}}{2}
     \hat{U}^{n} + \hat{R}^{n} \:\: , \:\:\: \hat{R}^{n} \equiv
     \hat{P}^{(1)n}- \frac{\beta_{1}}{2\beta_{0}} \hat{P}^{(0)n} \:\: .
  \end{equation}
In order to determine $ \hat{U} $, we expand the matrix (10)
following ref.\ \cite{FP}:
  \begin{equation}
    \hat{P}^{(0)n} = \sum_{i=1}^{3} \hat{e}_{i}^{n} \lambda _{i}^{n}
  \end{equation}
with  $ \lambda _{i}^{n} $ denoting the eigenvalues of $ \hat{P}^{(0)n}
$. The arbitrarily normalized eigenvectors $ \vec{v}_{i} $
  \begin{equation}
    \vec{v}_{i} =  \left( \begin{array}{c}
       2fP_{qG}^{(0)}/( \lambda _{i}-P_{qq}^{(0)})      \\[0.1cm]
                         1                              \\[0.1cm]
       6P_{qG}^{(0)}/ ( \lambda _{i}-\frac{9}{4}P_{qq}^{(0)})
			   \end{array} \right)
  \end{equation}
build up the projection matrices
  \begin{equation} (\hat{e}_{i})_{\alpha \beta} = v_{\alpha i}
                   v^{-1}_{i \beta } \:\: ,
  \end{equation}
where $ v_{\alpha i} \equiv (\vec{v}_{i})_{\alpha} $. Substituting
$ \hat{U} = \sum_{i,j=1}^{3} \hat{e}_{i} \hat{U} \hat{e}_{j} $ and (14)
into eq.\ (13) one readily derives
  \begin{equation}
    \hat{U} = -\frac{2}{\beta _{0}} \sum_{i=1}^{3} \hat{e}_{i} \hat{R}
    \hat{e}_{i} + \sum_{\stackrel{i,j=1}{i\ne j}}^{3} \frac{\hat{e}_{i}
    \hat{R} \hat{e}_{j}}
    {\lambda _{j}-\lambda _{i}-\beta _{0}/2} \:\: .
  \end{equation}
For simplicity of writing, we have suppressed the superscript $n$ in
(15--17). Eqs.\ (12) and (17) then yield the explicit singlet solution
\begin{eqnarray}
  \vec{q}^{\,n}(Q^{2}) &=&  \left\{ \sum_{i=1}^{3}
    \left( \frac{\alpha_{S}(Q^{2})}{\alpha_{S} (Q_{0}^{2})} \right)
     ^{-2 \lambda_{i}^{n} /\beta_{0}} \left[ \hat{e}_{i}^{n}
     + \frac{\alpha_{S}(Q_{0}^{2})-\alpha_{S}(Q^{2})} {2\pi} \frac{2}
     {\beta_{0}} \hat{e}_{i}^{n}\hat{R}^{n}\hat{e}_{i}^{n} \right.
                                                       \right.  \\
    & & \mbox{} \!\!\!\!\!\! \left. \left. - \sum_{j \ne i} \left(
    \frac{\alpha_{S} (Q_{0}^{2})}{2\pi} - \frac{\alpha_{S}(Q^{2})}
     {2\pi} \left( \frac{\alpha_{S}(Q^{2})} {\alpha_{S}(Q_{0}^{2})}
     \right) ^{-2 (\lambda_{j}^{n}-\lambda_{i}^{n}) /\beta_{0}} \right)
     \frac{\hat{e}_{i}^{n}\hat{R}^{n}\hat{e}_{j}^{n}}
     {\lambda_{j}^{n}-\lambda_{i}^{n}-\beta_{0}/2}
     \right] \right\} \vec{q}^{\,n}(Q_{0}^{2}) \:\: .     \nonumber
\end{eqnarray}

Finally, the individual quark distributions can be obtained from the
singlet and non-singlet moments (11) and (18) using the
relation
  \begin{equation}
     q_{i}^{n}(Q^{2}) + \bar{q}_{i}^{n}(Q^{2}) = \frac{1}{f}
      \Sigma^{n}(Q^{2}) - \frac{1}{i} v_{+,i^{2}-1}^{n}(Q^{2})
    + \sum_{k=i+1}^{f} \frac{1}{k(k-1)} v_{+,k^{2}-1}^{n}(Q^{2}) \:\: .
  \end{equation}
\section{Phenomenological Analysis}
We begin with the question whether or not a very light gluino is
consistent with the high-statistics fixed-target DIS measurements. To
find the answer it is not necessary to carry out a complete global fit
to all data available.
Instead we follow ref.\ \cite{GRV2} and proceed in two steps. First,
we study the valence quark region and the strong coupling
constant. Using the result as an input we then investigate the more
interesting regime at smaller $x$ where the gluon and sea
distributions play an important role. Only this second part of the
analysis requires new fits to data.

For the valence quark description of $F_{2}$ and $xF_{3}$ in standard
QCD we adopt the NLO proton distributions and the QCD scale
parameter from the global fit KMRS(B$_{-}$) \cite{KMRS}\footnote{In
refs.\ \cite{KMRS} and \cite{CCFR} a slightly different NLO relation
between $ \alpha_{S} $ and $ \Lambda $ has been used. We have
transformed their results to the convention (2) employed here,
resulting in a 10 MeV increase of $ \Lambda ^{(4)}$.}. At the
reference scale $ Q_{0}^{2} = 4 \mbox{ GeV}^{2} $, they are given by
\begin{eqnarray}
  x(u_{v}+d_{v})(x,Q_{0}^{2}) &=& 0.385 \, x^{0.27} (1 +  9.9 \sqrt{x}
     + 17.7x) (1-x)^{3.93}                               \nonumber  \\
  xd_{v}(x,Q_{0}^{2})         &=&  1.50 \, x^{0.61} (1 + 1.08 \sqrt{x})
    (1-x)^{4.68}
\end{eqnarray}
and $ \Lambda^{(4)} = 200 \mbox{ MeV} $. More recent fits exist,
the differences in the valence quarks being, however, marginal
and completely immaterial for the present purpose. Also the value of
$ \Lambda^{(4) }$ is consistent with more recent determinations,
for example from newer CCFR data giving $ \Lambda^{(4)} =
(220 \pm 50) \mbox{ MeV} $ \cite{CCFR}$^{2}$. With a gluino in the mass
range $ m_{\tilde{g}} < Q_{0} $ included it is still possible to
reproduce $ xF_{3}(x,Q^{2}) $ and the valence contribution
to $ F_{2}(x,Q^{2}) $ in the measured range, $ 5 \mbox{ GeV}^{2}
\stackrel{<}{{\scriptstyle \sim}} Q^{2} \stackrel{<}{{\scriptstyle
\sim}} 300 \mbox{ GeV}^{2} $ and $ 0.01 \stackrel{<}{{\scriptstyle
\sim}} x \stackrel{<}{{\scriptstyle \sim}} 0.7 $, within less than
$ 0.5 \, \% $ by a slight modification of the valence quark
input and a suitable choice of $ \Lambda^{(4)}_{\tilde{g}} $.
Hence no new valence fit to data is necessary. Optimization of this
agreement leads to
\begin{eqnarray}
  x(u_{v}+d_{v})(x,Q_{0}^{2})_{\tilde{g}}
      &=& 0.388 \, x^{0.27} (1 +  9.9 \sqrt{x} + 17.6x) (1-x)^{3.975}
						     \nonumber  \\
  xd_{v}(x,Q_{0}^{2})_{\tilde{g}}
      &=&  1.51 \, x^{0.61} (1 + 1.09 \sqrt{x}) (1-x)^{4.74}
\end{eqnarray}
together with $ \Lambda^{(4)}_{\tilde{g}} = 30 \mbox{ MeV} $.

The strong coupling constants $ \alpha_{S}(Q) $ in the two scenarios
(20) and (21) coincide at $ Q^{2} = 12 \mbox{ GeV}^{2} $. The
continuity of $ \alpha_{S} $ at the heavy quark thresholds
$ Q^{2} = m_{h}^{2} $ ($ h=c,b $) then implies for the NLO scale
parameters $ \Lambda^{(f)} $ in the $\overline{\mbox{MS}} $ scheme:
\begin{eqnarray}
  \Lambda^{(3,4,5)}      &=& 248, 200, 131  \mbox{ MeV}  \nonumber \\
  \Lambda_{\tilde{g}}^{(3,4,5)} &=& 56.5, 30, 10.6 \mbox{ MeV} \:\: ,
\end{eqnarray}
where $ m_{c} = 1.5 \mbox{ GeV} $, $ m_{b} = 4.5 \mbox{ GeV} $
and $ m_{\tilde{g}} < m_{c} $ has been used. Obviously, these
values of $ \Lambda_{\tilde{g}}^{(f)} $ are independent of the
gluino mass, but apply only above the gluino threshold $ Q =
m_{\tilde{g}} $. The gluino mass dependent values of
$ \Lambda_{\tilde{g}}^{(3)} $
relevant below the gluino threshold will be given later.  Using
eq.\ (22) we find $ \alpha_{S}(M_{Z}) = 0.128 \: (0.109) $  with
(without) the gluino. The result in brackets agrees with the recent
CCFR value, $ \alpha_{S}(M_{Z}) = 0.111 \pm 0.005 $ \cite{CCFR}
obtained from a standard QCD analysis. A similar uncertainty
in $ \alpha_{S}(M_{Z}) $ is expected for the gluino case. The
existence of a light gluino would, however, not only modify the
DIS results on $ \alpha_{S} (Q) $ and the extrapolation to $ Q= M_{Z} $
but also change the direct LEP determinations of $ \alpha_{S}(M_{Z}) $
due to contributions to virtual and real radiative corrections
\cite{ENR}. With (without) a light gluino  one obtains
\cite{ENR,Beth2} $ \alpha_{S}(M_{Z}) = 0.132 \: (0.120) \pm 0.006 $
from hadronic event shapes and $ \alpha_{S}(M_{Z}) = 0.124 \: (0.122)
\pm 0.009 $ from the hadronic width $ \Gamma _{h} $ of the $ Z^{0} $.
Note that the original results of ref.\ \cite{ENR} have been updated
with respect to the latest value of $ \Gamma _{h} $.  Cleary, within
present uncertainties, both scenarios give a consistent picture.

We now turn to the more interesting part of the analysis dealing with
the gluon and sea densities. For this, we adopt the radiative approach
developed in refs.\ \cite{GRV1,GRV2,GRVp,GRVg}. This approach provides
a unified description of hadronic and photonic parton distributions
which agrees with experiment, and which predicted the steep rise of the
structure function $ F_{2} $ at small $x$ discovered recently at HERA
\cite{ZEUS,H1}. The charactistic features of this framework are a very
low scale $ (\mu / \Lambda ^{(3)})^{2} \simeq 5 \div 6 $ at which the
Altarelli-Parisi evolution is started and valence-like input
densities.
For light gluinos with $ {m}_{\tilde{g}} > \mu $, this procedure has
the advantage that the gluino can approximately be treated like a
massive quark with the boundary condition (6). It is generated
similarly to the strange sea discussed below. In this way one can avoid
the introduction of a gluino input distribution except for almost
massless gluinos with $ m_{\tilde{g}} < \mu $. For the gluon and sea
quark inputs we use \cite{GRV2}
\begin{eqnarray}
  xG(x,\mu^{2}) &=& A x^{\alpha} (1-x)^{\beta}          \nonumber  \\
  x\bar{u}(x,\mu^{2}) &=& x\bar{d}(x,\mu^{2}) = A^{\prime}
     x^{\alpha^{\prime}} (1-x)^{\beta^{\prime}}                    \\
  xs(x,\mu^{2}) &=& x\bar{s}(x,\mu^{2}) = 0 \:\: .      \nonumber
\end{eqnarray}
The vanishing of the strange sea at $ \mu^{2} $ leads to SU(3) breaking
in the sea in accordance with experimental results from opposite-sign
dimuon events in neutrino-nucleon scattering \cite{CCFRs}.  The value
of $\mu $ is fixed by the momentum fraction $ <\! x\! >_{v}(\mu^{2}) $
carried by the valence quarks. For the standard QCD case we take $
\mu^{2} = 0.3 \mbox{ GeV}^{2} $ corresponding to $ <\! x\! >_{v} \simeq
0.6 $ \cite{GRV2}. Keeping this fraction also in the presence of a
gluino, the scale $\mu $ decreases with the gluino mass due to the
slower evolution at low $ Q^{2} $. This correlation is shown in table 1
for typical gluino masses. In order to illustrate the sensitivity
of the outcome of the subsequent fits to the exact value of $\mu $ we
also give results for $ \mu^{2} = 0.36 \mbox{ GeV}^{2} $, that is, for
a 10\% increase of $ \mu $.

The input parameters appearing in (23) are fitted to the available
high-statistics data on $ F_{2}^{\, ed} $ and $ F_{2}^{\, \mu d} $
from SLAC \cite{SLAC}, BCDMS \cite{BCDMS}\footnote{We take the
BCDMS data as used in \cite{MV}, i.e.\ with a shift of the
central values due to the main systematic error. However, in the
$x$-region considered here the effect is not essential.} and NMC
\cite{NMC} in the region of good sensitivity, i.e.\ at $ x < 0.3 $.
Statistical and systematical errors are added in quadrature. As in
ref.\ \cite{MRS} the BCDMS data is normalized down by 2\% relative to
the SLAC results. In order to suppress possible higher twist effects
we restrict our fits to measurements at $ Q^{2} > 5 \mbox{ GeV}^{2} $.
An additional cut on the invariant mass $W$ of the hadronic final
state is not neccessary in view of the low $x$-values involved.
In total, we are left with 147 data points.
Additional high precision data on $ F_{2}^{\, \nu N} $, $ F_{2}^{\, p}
$ and the ratio $ F_{2}^{\, n} / F_{2}^{\, p} $ mainly influence the
flavour decomposition, and are unimportant for the gluino issue.
Moreover, by using the valence distributions from ref.\ \cite{KMRS}
the main features of these measurements are incorporated.
Likewise, data on Drell-Yan lepton pair production mainly constrain
the  shape of the sea quark densities at large $x$ where they are
small and not relevant for the present study.
On the other hand, data on direct photon production in proton-proton
collisions \cite{WA70} may be relevant, since this process directly
probes the gluon distribution which is most affected by light
gluinos. However, unlike in the case of DIS and the Drell-Yan process,
the hard photon production cross sections in NLO do get modified in the
presence of gluinos.
These modifications have not yet been calculated. Hence, we disregard
the direct photon data in the fits for theoretical consistency but take
them into account approximately as discussed below.

\begin{table}[ht]
  \begin{center}
  \begin{tabular}{|c||c|c|c||c|c|c||c|}\hline
	 &     &       &        &        &       &         & \\[-0.4cm]
  $ m_{\tilde{g}}          $ & $\Lambda^{(3)}/\mbox{MeV} $ &
  $ \mu^{2}/\mbox{GeV}^{2} $ & $ <\! x\! >_{v}(\mu^{2})  $ &
  $ \alpha^{\prime}        $ & $ \beta^{\prime}          $ &
  $ A^{\prime}             $ & $ \chi^{2}                $   \\[0.1cm]
						   \hline \hline
	 &     &       &        &        &       &        &  \\[-0.4cm]
    0.4  & 118 & 0.09  & 0.617  & 0.490  & 8.03  & 0.138  &  140  \\
    0.7  & 148 & 0.125 & 0.618  & 0.519  & 8.02  & 0.134  &  139  \\
    1.0  & 168 & 0.155 & 0.615  & 0.514  & 8.03  & 0.131  &  137  \\
    1.3  & 184 & 0.18  & 0.615  & 0.525  & 8.06  & 0.130  &  137  \\
   no $\tilde{g} $
	 & 248 & 0.30  & 0.614  & 0.529  & 7.93  & 0.127  &  134
				   	               \\[0.1cm] \hline
	 &     &       &        &        &       &        &  \\[-0.4cm]
    0.4  & 118 & 0.12  & 0.572  & 0.362  & 7.66  & 0.134  &  134  \\
    0.7  & 148 & 0.155 & 0.578  & 0.371  & 7.63  & 0.130  &  134  \\
    1.0  & 168 & 0.19  & 0.576  & 0.364  & 7.59  & 0.129  &  133  \\
    1.3  & 184 & 0.22  & 0.575  & 0.364  & 7.56  & 0.128  &  132  \\
   no $\tilde{g} $
         & 248 & 0.36  & 0.574  & 0.367  & 7.44  & 0.125  &  130
			                 	       \\[0.1cm] \hline
  \end{tabular}
  \caption{Fits of DIS structure functions to experimental data on
	   $ F_{2}^{\, d} $ at $ x < 0.3 $ and $ Q^{2} > 5 \mbox{ GeV}
	   ^{2} $ (147 data points) in standard QCD (no $ \tilde{g} $)
	   and including a light gluino. The input distributions
           are given in eq.\ (23) and specified further in the text.}
  \end{center}
\end{table}
Since the gluon normalization $A$ is fixed by the momentum sum rule,
we have five free parameters describing the input distributions (23).
These are fitted to the data selected above in the standard QCD case
and for several gluino masses $ m_{\tilde{g}} $ between $ 0.4
\mbox{ GeV} $ and $ 1.3 \mbox{ GeV} $.
The results of the fits are given in table 1. The outcome is almost
insensitive to variations of the two shape variables of the
gluon density in the range of values favoured by direct photon
measurements \cite{WA70,ABFOW}. For this reason, we simply take
\begin{equation}
  \alpha = 2.0 \: (1.5) \:\: , \:\:\: \beta = 4.0
\end{equation}
in the fits summarized in the upper (lower) part of the table.
Also the sea quark shape parameters $ \alpha^{\prime} $ and
$ \beta^{\prime} $ are only slightly affected by the inclusion of
a light gluino. However, the overall normalization $ A^{\prime} $
shows a small but significant rise with decreasing gluino mass.
This is exactly what one expects physically since the gluon density,
and hence the evolution of quark-antiquark pairs, is somewhat
diminished due to gluino radiation.
The quality of the above fits is as good as the one of the recent
MRS global analysis \cite{MRS}, if one compares the description
of the data used in both fits. As can be seen from the $ \chi^{2} $
values in the last row of table 1 there is no evidence from the
fits in favour or against gluinos with masses below $ 1 \mbox{ GeV} $.
This conclusion agrees with ref.\ \cite{BB}.
\section{Resulting Parton Distributions}
Using the results from the fits outlined in the previous section, we
now discuss the parton densities of the nucleon and the structure
function $ F_{2}^{\, p} $. This will also enable us to answer the second
question raised in the introduction concerning the potential of future
DIS measurements at HERA in testing the existence of a very light
gluino. As reference cases for comparison we focus on the standard QCD
fit with $ \mu^{2} = 0.30 \mbox{ GeV}^{2} $ and the gluino scenario
with $ m_{\tilde{g}} = 0.7 \mbox{ GeV} $ and $ \mu^{2} = 0.125
\mbox{ GeV}^{2} $.

It is instructive to consider briefly the fractions $ <\! x\! >_{f}
= \int_{0}^{1} \! dx \, xf(x,Q^{2}) $ of the nucleon momentum
carried by quarks, gluons and gluinos. Their $ Q^{2} $-evolution is
shown in fig.\ 1. As the main effect, the average gluon momentum is
lowered by gluino radiation. This reduction is fed back to the
quarks at large $ Q^{2} $. Below threshold, at $ Q^{2} \stackrel{<}
{{\scriptstyle \sim}} m_{\tilde{g}}^{2} $, the change of the quark and
gluon momentum fractions reflects the slower running of $ \alpha_{S}
(Q^{2}) $ in the presence of a gluino.  Asymptotically, the gluino
momentum reaches about 10\% of the proton momentum.  Moreover, one
can see that already at $  Q^{2} \simeq 10  \mbox{ GeV}^{2} $ the
gluino component gives a non-negligible contribution to the momentum
sum rule.

Figs.\ 2 and 3 exhibit the sea quark distributions $ x\bar{u} = x\bar{d}
$ and the gluon density $ xG $ at $ Q^{2} = 10 \mbox{ GeV}^{2} $. While
the inclusion of the gluino has virtually no effect on the sea quarks,
the gluon distribution is reduced by about 15\% over the range $ x
\stackrel{>}{{\scriptstyle \sim}} 0.05 $. This shrinkage is, however,
not expected to lead to phenomenological problems, most notably, in
direct photon production, given the present theoretical and
experimental uncertainties and taking into account the partial
compensation by the larger value of $ \alpha_{S}(Q^{2}) $ at high $
Q^{2} $.

The predictions in the small-$x$ region and the $ Q^{2} $-evolution
of $ x\bar{u} $ and $ xG $ are depicted in figs.\ 4 and 5.  Also here,
the gluino effect on the sea quark density can be neglegted for all
practical purposes. Furthermore, the depletion of the gluon distribution
shown in fig.\ 3 for large $x$ disappears as $x$ decreases. All
effects depend only weakly on $ Q^{2} $. In particular, in the region
of $x$ and $ Q^{2} $ currently under investigation at HERA, it appears
impossible to uncover a light gluino in the structure
functions.
This conclusion is further corroborated in fig.\ 6, where $ F_{2}^{\, p}
$ is displayed in this kinematical region together with first HERA data
\cite{ZEUS,H1}. The tiny down-shifts of $ F_{2}^{\, p} $ are mainly due
to a smaller charm contribution\footnote{Above threshold the charm
quark is treated as a massless flavour. Using instead the complete
photon-gluon fusion cross section  for the charm contribution results
in a decrease of $ F_{2}^{\, p} $ by $ 5 \div 10\, \% $ at small $x$.
For a detailed discussion see \cite{GRS}.} which in turn is a
consequence of the lower gluon density. The standard QCD results with
$ \mu^{2} = 0.3 \mbox{ GeV}^{2} $ are practically identical to the q
pre-HERA predictions of refs.\ \cite{GRV2}. The uncertainty in these
predictions is illustrated by the expectations for $ \mu^{2} = 0.36
\mbox{ GeV}^{2} $.

The above findings differ from the results obtained in ref.\ \cite{BB}.
While, according to this paper, at $ Q^{2} = 10 \mbox{ GeV}^{2} $ the
quark and gluon distributions are only affected at the $ 10\% $ level,
at $ Q^{2} = 100 \mbox{ GeV}^{2} $ and $ 10^{-4} \stackrel{<}
{{\scriptstyle \sim}} x \stackrel{<}{{\scriptstyle \sim}} 10^{-3} $
the presence of a light gluino gives rise to an increase of $ xu $ and
$ xG $ by about 50\%. Moreover, this strong evolution is roughly
independent of the gluino mass for $ m_{\tilde{g}} \stackrel{<}
{{\scriptstyle \sim}} 5 \mbox{ GeV} $. This disagrees also with ref.\
\cite{RS}, where a gluino of 5 GeV is shown to change $ F_{2}^{\, p} $
by at most a few per cent.

Turning to the $Q^{2}$-evolution of structure functions at large $x$,
in fig.\ 7 we illustrate the changes in $ F_{2}^{\, p} $ in the
region $ Q^{2}/x < s \simeq 10^{5} \mbox{ GeV}^{2} $ which will be
probed at HERA. The deviations from the standard QCD evolution are
somewhat bigger than the effects found in ref.\ \cite{RS} for
$ m_{\tilde{g}} = 5 \mbox{ GeV} $. This is expected because of the
lower gluino mass considered here. However, also for such light
gluinos the deviations do not exceed a few per cent and will therefore
be very hard to detect \cite{BKIR}, in contrast to the expectation in
ref.\ \cite{ENR} based on the enhancemant of $ \alpha _{S} $ at high
scales. This conclusion suggests to return once more to the gluino
content in the nucleon considered at the beginning of this section
which may play a role in  more direct searches at $ep $- and hadron
colliders.

Fig.\ 8 shows distributions in $x$ and the evolution of the gluino
component together with the strange sea also generated radiatively
from a vanishing input. The threshold dependence is illustrated for
the gluino by considering three different masses $ m_{\tilde{g}} = 0.4
$, 0.7  and $ 1.0  \mbox{ GeV} $. At large $x$, $ \tilde{g}(x,Q^{2}) $
is about three times bigger than $ s(x,Q^{2}) $ and, incidentally,
rather similar to $ \bar{u}(x,Q^{2}) $. The ratio $ \tilde{g}/s $ rises
with decreasing $x$, reaching a value around five at $ x \simeq 10^{-4}
$ in accordance with the expectation from the differences in the colour
factors of the relevant splitting functions in eq.\ (10). The relative
abundance of gluinos depends only weakly on $ Q^{2} $ except close to
the threshold.
Comparing these predictions with the gluon and gluino distributions
given in ref.\ \cite{BB} we find rough agreement at $ Q^{2} = 10
\mbox{ GeV}^{2} $. However, at $ Q^{2} = 100 \mbox{ GeV}^{2} $ and
small $x$ a gluino with $ m_{\tilde{g}} \approx 0 $ is predicted to be
about three times as abundant as what we find. Moreover, it dominates
the corresponding up-quark density by almost a factor of nine in
contrast to the expectation from eq.\ (10). We have checked that
our evolution program reproduces the results of ref.\ \cite{RS} if the
same input is used. Therefore, we believe that fig.\ 8 represents a
reasonable estimate of the gluino content of the
proton, if a light gluino exists.
\section{Summary}
We have investigated the effects of a very light gluino with
$ m_{\tilde{g}} \stackrel{<}{{\scriptstyle \sim}} 1 \mbox{ GeV} $
on the quark and gluon distributions of the nucleon and the DIS
structure functions. The whole range in Bjorken-$x$ and $ Q^{2} $
which is being probed in fixed-target experiments and at HERA has been
taken into consideration. We have adopted the framework described in
refs.\ \cite{GRV1,GRV2} in which the parton densities are generated
radiatively from valence-like distributions at a very low input scale.
This procedure leads to rather definite predictions at small $x$ which
are in agreement with first HERA results \cite{ZEUS,H1}. Moreover, in
the present application it has the advantage that the light gluino can
be included in the $ Q^{2} $-evolution similarly as a massive quark,
i.e.\ without introducing a perturbatively uncalculable and
phenomenologically rather unconstrained gluino input function.

Using $ \Lambda^{(f)}_{\overline{MS}} $ as determined from a valence
quark analysis of $F_{2} $ and $ xF_{3} $ in the presence of a light
gluino, the valence-like input distributions have been fitted to the
relevant high-statistics data on $ F_{2}^{\, d} $.
Globally, we find that the gluino carries about $ 5 \div 10 \,
\% $ of the nucleon momentum. On the other hand, the effects on the
quark and gluon distributions at $ x \stackrel{<}{{\scriptstyle \sim}}
10^{-2} $ are completely negligible, even after evolution over a few
orders of magnitude in $ Q^{2} $. The fits to fixed-target data
including a light gluino are almost as good as the standard QCD fits.
Moreover, we have shown that the structure function $F_{2}^{\, p}
$ is practically insensitive to light gluinos also in the new
kinematical region covered by HERA. We thus conclude that measurements
of DIS structure functions are not able to discriminate between the
presence and absence of light gluinos.

Then, in order to test the light-gluino hypothesis one has to study
final state signatures such as jet rates and and displaced vertices.
In $ \bar{p}p $, $ pp $ and $ ep $ collisions,
this involves the gluino density in the proton. We have presented a
radiative estimate for this component assuming that the gluino
distribution vanishes at the threshold $  Q^{2} = m_{\tilde{g}}^{2} $.
Far from threshold the gluino density is predicted to be about three
to five times higher than the strange quark density.

\pagebreak

\begin{figure}[1]
\centerline{
\begin{picture}(400,330)(0,-20)
\put(0,10){\strut\epsffile{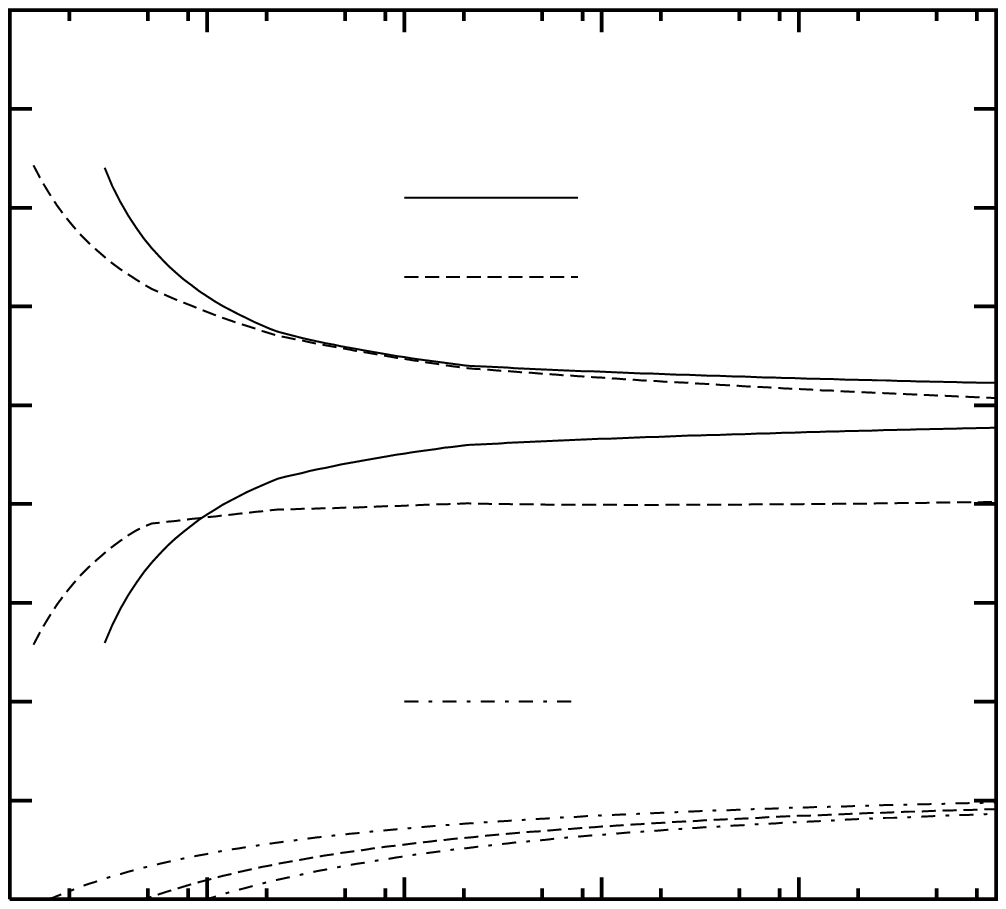}}
\Text(100,255)[l]{{\LARGE $ <\! x\! >_{f} $}}
\Text(178,-10)[l]{{\Large $Q^{\, 2}(\mbox{GeV}^{\, 2})$}}
\Text(242,233)[l]{{\large no $ \tilde{g} $}}
\Text(242,210)[l]{{\large $ m_{\tilde{g}} = 0.7 \mbox{ GeV}$}}
\Text(242,88)[l]{{\large $ m_{\tilde{g}} = \!
 \begin{array}{c} {\scriptstyle 0.4} \\[-0.2cm] {\scriptstyle 1.0}
  \end{array} \! \mbox{ GeV} $}}
\Text(125,220)[c]{{\large $ \Sigma $}}
\Text(125,157)[c]{{\large $ G $}}
\Text(125,58)[c]{{\large $ \tilde{g} $}}
\Text(71,15)[c]{{\Large $10^{-1}$}}
\Text(124,15)[c]{{\Large $1$}}
\Text(181,15)[c]{{\Large $10$}}
\Text(239,15)[c]{{\Large $10^{\, 2}$}}
\Text(295,15)[c]{{\Large $10^{\, 3}$}}
\Text(352,15)[c]{{\Large $10^{\, 4}$}}
\Text(36,32)[l]{{\Large $0.0$}}
\Text(36,89)[l]{{\Large $0.2$}}
\Text(36,146)[l]{{\Large $0.4$}}
\Text(36,203)[l]{{\Large $0.6$}}
\Text(36,260)[l]{{\Large $0.8$}}
\end{picture}
}
\caption{The NLO $ Q^{2} $-evolution of the momentum fractions
$ <\! x\! >_{f} $ carried by quarks and gluons in standard QCD
and including a light gluino with $ m_{\tilde{g}} = 0.7 \mbox{ GeV} $.
The gluino momentum fraction is given for varying gluino masses.}
\end{figure}

\pagebreak

\begin{figure}[2]
\centerline{
\begin{picture}(400,400)(0,-10)
\put(0,10){\strut\epsffile{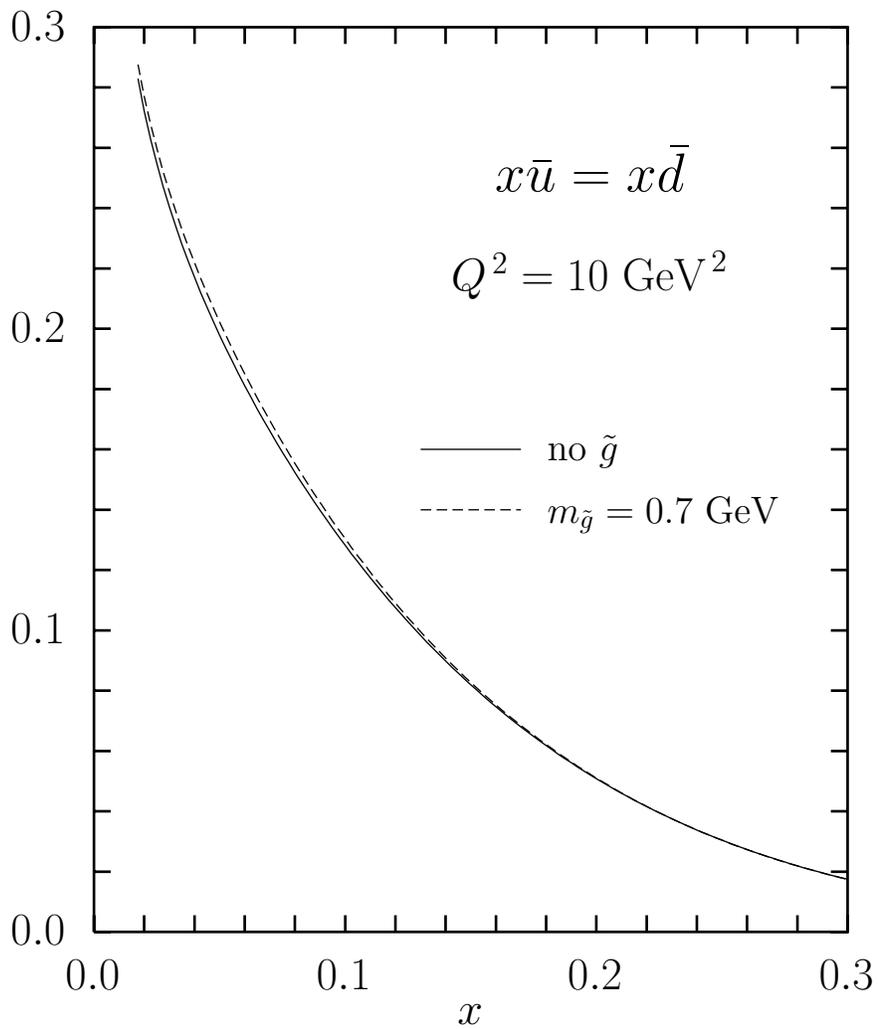}}
\Text(255,320)[c]{{\LARGE  $x\bar{u} = x\bar{d}$}}
\Text(255,280)[c]{{\Large  $Q^{\, 2} = 10 \mbox{ GeV}^{\, 2}$}}
\Text(239,213)[l]{{\large no $ \tilde{g} $}}
\Text(239,189)[l]{{\large $ m_{\tilde{g}} = 0.7 \mbox{ GeV}$}}
\Text(205,0)[l]{{\Large $x$}}
\Text(67,15)[c]{{\Large $0.0$}}
\Text(162,15)[c]{{\Large $0.1$}}
\Text(257,15)[c]{{\Large $0.2$}}
\Text(352,15)[c]{{\Large $0.3$}}
\Text(36,32)[l]{{\Large $0.0$}}
\Text(36,146)[l]{{\Large $0.1$}}
\Text(36,260)[l]{{\Large $0.2$}}
\Text(36,374)[l]{{\Large $0.3$}}
\end{picture}
}
\caption{NLO sea quark distribution  with and without a light gluino
in the $x$-region probed by fixed-target DIS.}
\end{figure}

\pagebreak

\begin{figure}[3]
\centerline{
\begin{picture}(400,400)(0,-10)
\put(0,10){\strut\epsffile{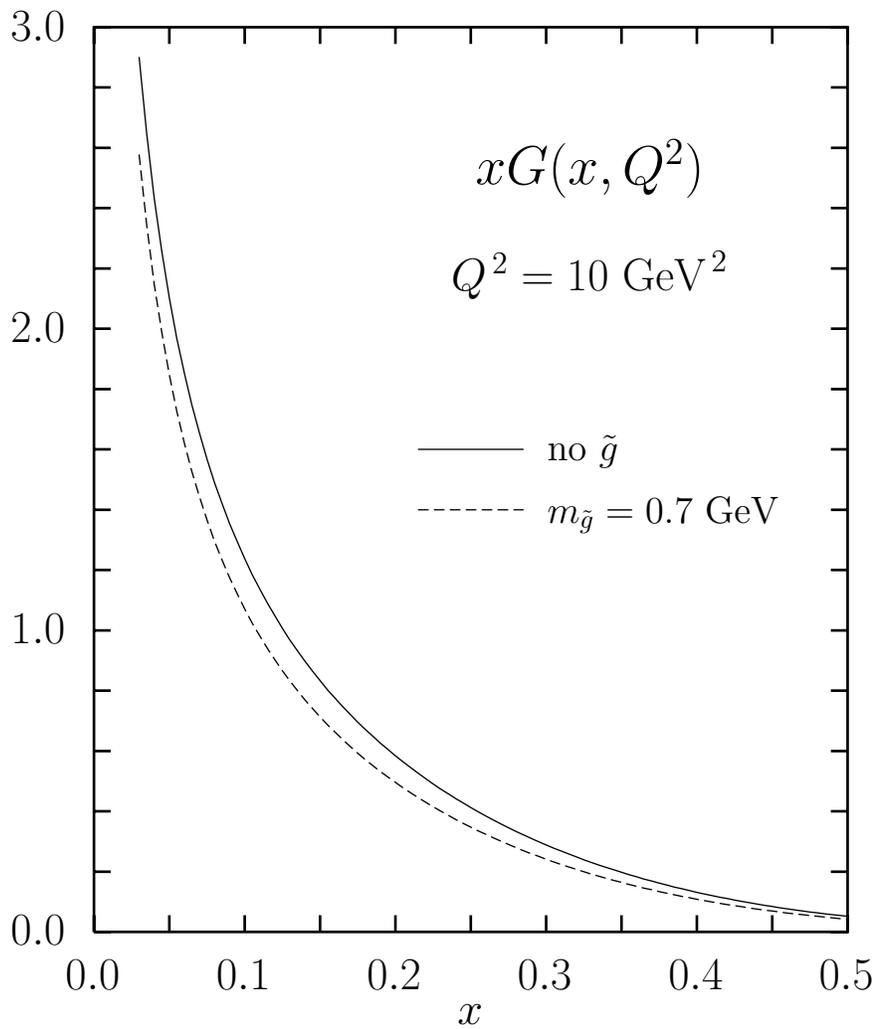}}
\Text(255,320)[c]{{\LARGE  $xG(x,Q^{2}$)}}
\Text(255,280)[c]{{\Large  $Q^{\, 2} = 10 \mbox{ GeV}^{\, 2}$}}
\Text(239,213)[l]{{\large no $ \tilde{g} $}}
\Text(239,189)[l]{{\large $ m_{\tilde{g}} = 0.7 \mbox{ GeV}$}}
\Text(205,0)[l]{{\Large $x$}}
\Text(67,15)[c]{{\Large $0.0$}}
\Text(124,15)[c]{{\Large $0.1$}}
\Text(181,15)[c]{{\Large $0.2$}}
\Text(238,15)[c]{{\Large $0.3$}}
\Text(295,15)[c]{{\Large $0.4$}}
\Text(352,15)[c]{{\Large $0.5$}}
\Text(36,32)[l]{{\Large $0.0$}}
\Text(36,146)[l]{{\Large $1.0$}}
\Text(36,260)[l]{{\Large $2.0$}}
\Text(36,374)[l]{{\Large $3.0$}}
\end{picture}
}
\caption{The NLO gluon distribution with and without a light gluino
in the $x$-range probed by fixed-target DIS.}
\end{figure}

\pagebreak

\begin{figure}[4]
\centerline{
\begin{picture}(400,420)(0,-10)
\put(0,10){\strut\epsffile{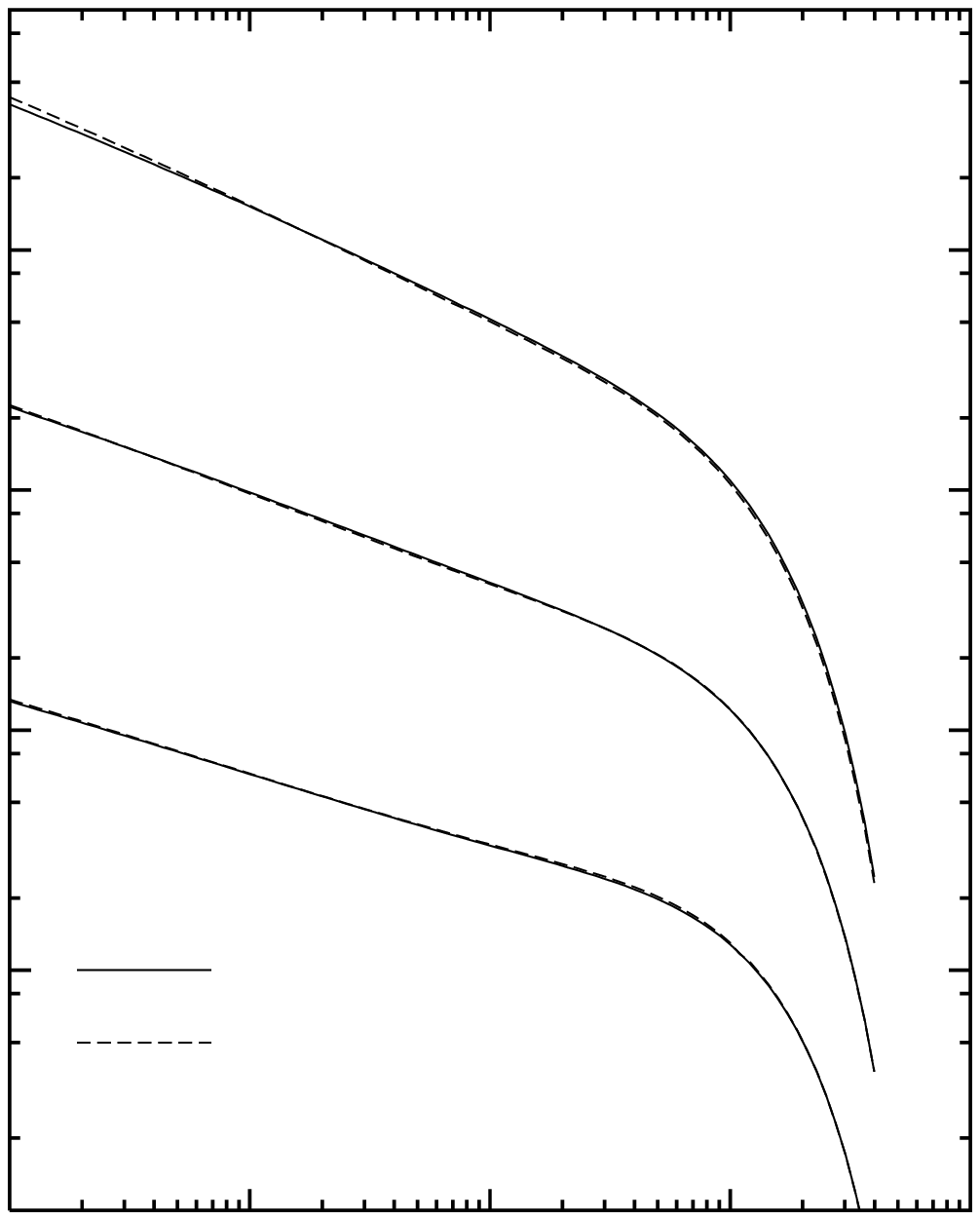}}
\Text(230,332)[l]{{\LARGE  $x\bar{u}(x,Q^2)$}}
\Text(135,102)[l]{{\large no $ \tilde{g} $}}
\Text(135,80)[l]{{\large $ m_{\tilde{g}} = 0.7 \mbox{ GeV}$}}
\Text(170,0)[l]{{\Large $x$}}
\Text(110,353)[l]{{$ 10^{\, 4} = Q^{2} (\mbox{GeV}^{2}) $}}
\Text(110,266)[l]{{$ 100 $}}
\Text(110,181)[l]{{$ 10 $}}
\Text(260,289)[c]{{$ (\times 10) $}}
\Text(260,142)[c]{{$ (\times 0.1) $}}
\Text(71,15)[c]{{\Large $10^{-4}$}}
\Text(142,15)[c]{{\Large $10^{-3}$}}
\Text(213,15)[c]{{\Large $10^{-2}$}}
\Text(285,15)[c]{{\Large $10^{-1}$}}
\Text(352,15)[c]{{\Large $1$}}
\Text(31,34)[l]{{\Large $10^{-3}$}}
\Text(31,104)[l]{{\Large $10^{-2}$}}
\Text(31,175)[l]{{\Large $10^{-1}$}}
\Text(31,245)[l]{{\Large $1$}}
\Text(31,316)[l]{{\Large $10$}}
\Text(31,389)[l]{{\Large $10^{\, 2}$}}
\end{picture}
}
\caption{The small-$x$ behaviour of the sea quark density
$ x\bar{u} = x\bar{d} $ and its evolution in $ Q^{2} $.  Compared
are predictions  with and without a light gluino.}
\end{figure}

\pagebreak

\begin{figure}[5]
\centerline{
\begin{picture}(400,420)(0,-10)
\put(0,10){\strut\epsffile{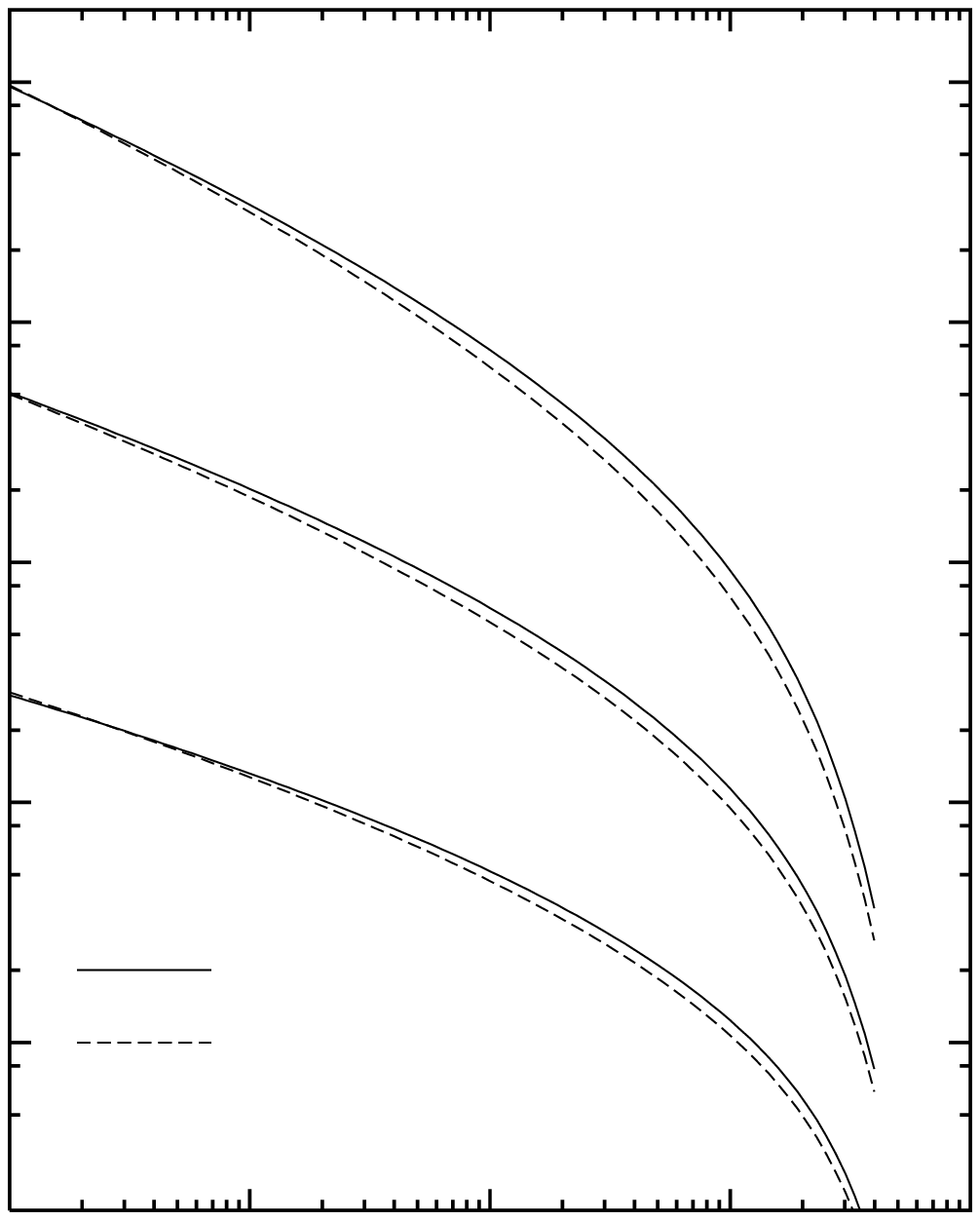}}
\Text(230,325)[l]{{\LARGE  $xG(x,Q^2)$}}
\Text(135,102)[l]{{\large no $ \tilde{g} $}}
\Text(135,80)[l]{{\large $ m_{\tilde{g}} = 0.7 \mbox{ GeV}$}}
\Text(170,0)[l]{{\Large $x$}}
\Text(110,356)[l]{{$ 10^{\, 4} = Q^{2} (\mbox{GeV}^{2}) $}}
\Text(110,269)[l]{{$ 100 $}}
\Text(110,183)[l]{{$ 10 $}}
\Text(260,270)[c]{{$ (\times 10) $}}
\Text(260,125)[c]{{$ (\times 0.1) $}}
\Text(71,15)[c]{{\Large $10^{-4}$}}
\Text(142,15)[c]{{\Large $10^{-3}$}}
\Text(213,15)[c]{{\Large $10^{-2}$}}
\Text(285,15)[c]{{\Large $10^{-1}$}}
\Text(352,15)[c]{{\Large $1$}}
\Text(31,83)[l]{{\Large $10^{-1}$}}
\Text(31,152)[l]{{\Large $1$}}
\Text(31,223)[l]{{\Large $10$}}
\Text(31,296)[l]{{\Large $10^{\, 2}$}}
\Text(31,367)[l]{{\Large $10^{\, 3}$}}
\end{picture}
}
\caption{Same as fig.\ 4, but for the gluon distribution.}
\end{figure}

\pagebreak

\begin{figure}[6]
\centerline{
\begin{picture}(400,540)(0,-10)
\put(0,10){\strut\epsffile{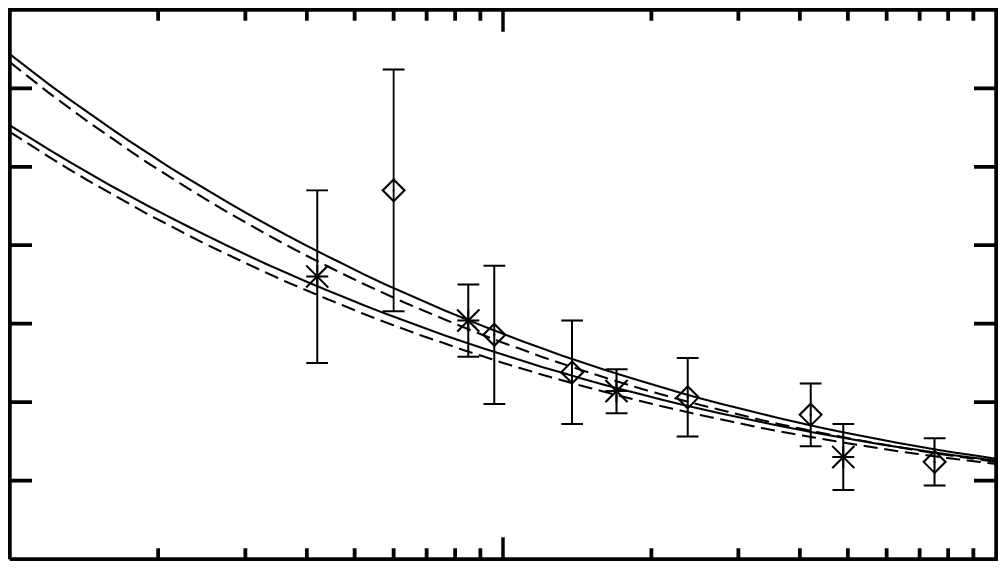}}
\put(0,169){\strut\epsffile{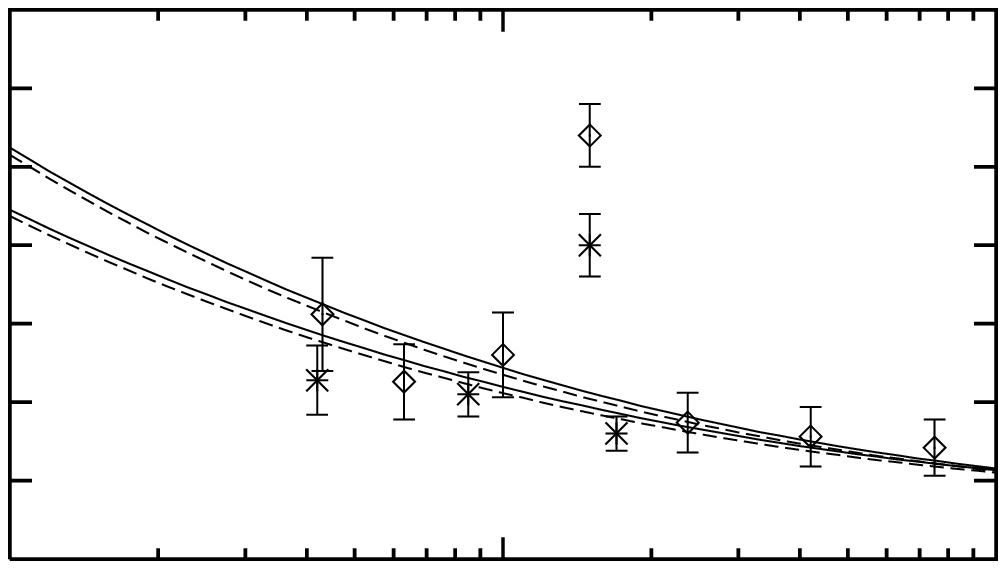}}
\put(0,328){\strut\epsffile{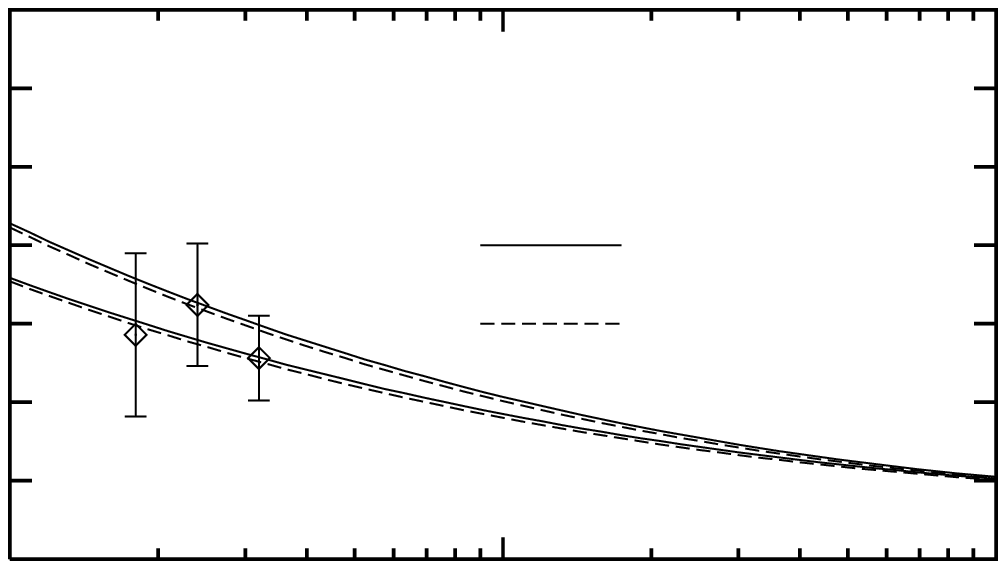}}
\Text(100,477)[l]{{\LARGE $ F_{2}^{\, p}(x,Q^{2}) $}}
\Text(253,441)[l]{{\large no $ \tilde{g} $}}
\Text(253,417)[l]{{\large $ m_{\tilde{g}} = 0.7 \mbox{ GeV}$}}
\Text(253,312)[l]{{\large H1}}
\Text(253,281)[l]{{\large ZEUS}}
\Text(275,0)[l]{{\Large $x$}}
\Text(90,55)[l]{{\large $ Q^{\, 2} = 30 \mbox{ GeV}^{\, 2} $}}
\Text(90,214)[l]{{\large $ Q^{\, 2} = 15 \mbox{ GeV}^{\, 2} $}}
\Text(90,373)[l]{{\large $ Q^{\, 2} = 8.5 \mbox{ GeV}^{\, 2} $}}
\Text(71,15)[c]{{\Large $10^{-4}$}}
\Text(213,15)[c]{{\Large $10^{-3}$}}
\Text(356,15)[c]{{\Large $10^{-2}$}}
\Text(50,32)[l]{{\Large $0$}}
\Text(50,77)[l]{{\Large $1$}}
\Text(50,122)[l]{{\Large $2$}}
\Text(50,167)[l]{{\Large $3$}}
\Text(50,191)[l]{{\Large $0$}}
\Text(50,236)[l]{{\Large $1$}}
\Text(50,281)[l]{{\Large $2$}}
\Text(50,326)[l]{{\Large $3$}}
\Text(50,350)[l]{{\Large $0$}}
\Text(50,395)[l]{{\Large $1$}}
\Text(50,440)[l]{{\Large $2$}}
\Text(50,485)[l]{{\Large $3$}}
\end{picture}
}
\caption{NLO predictions for the proton structure function
$ F_{2}^{\, p} $ at small $x$ in comparison to first HERA
data from the ZEUS \protect\cite{ZEUS} and H1 \protect\cite{H1}
collaborations. The solid curves represent the standard QCD results
for the input scales $ \mu^{2} = 0.3 \mbox { GeV}^{2} $ (upper) and
$ \mu^{2}= 0.36\mbox { GeV}^{2} $ (lower). The dashed curves show the
corresponding predictions with a light gluino included.}
\end{figure}

\pagebreak

\begin{figure}[7]
\centerline{
\begin{picture}(400,330)(0,-20)
\put(0,10){\strut\epsffile{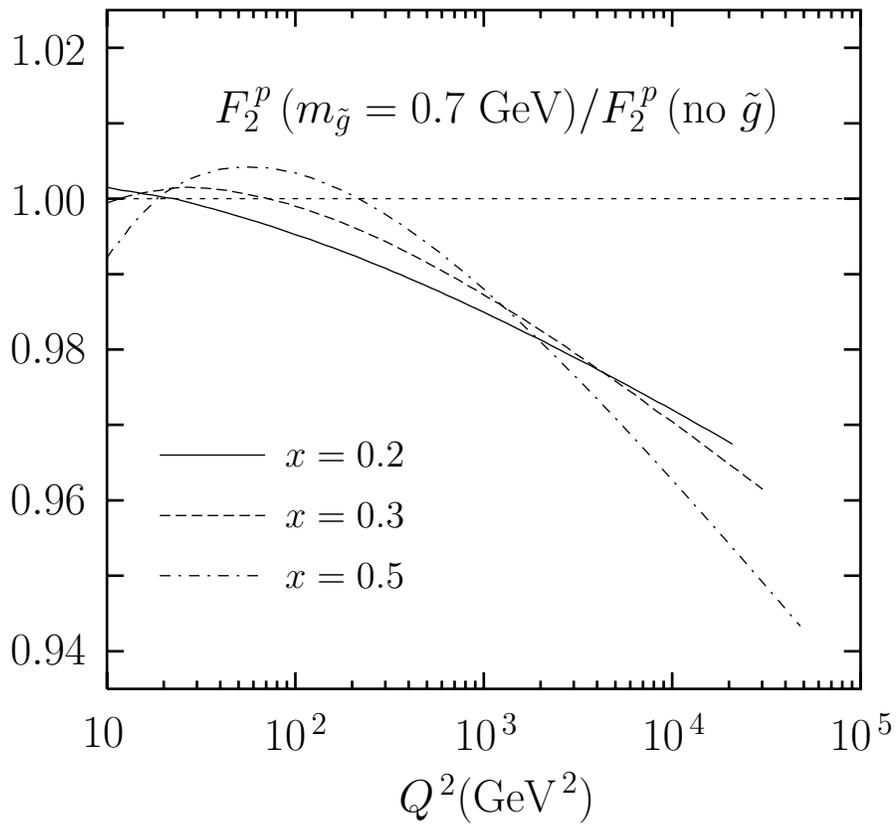}}
\Text(215,250)[c]{{\Large $ F_{2}^{\, p} \, (m_{\tilde{g}} = 0.7
	      \mbox{ GeV}) /  F_{2}^{\, p} \, (\mbox{no }\tilde{g})$}}
\Text(178,-10)[l]{{\Large $Q^{\, 2}(\mbox{GeV}^{\, 2})$}}
\Text(135,74)[l]{{\large $ x = 0.5 $}}
\Text(135,120)[l]{{\large $ x = 0.2 $}}
\Text(135,97)[l]{{\large $ x = 0.3 $}}
\Text(66,15)[c]{{\Large $10$}}
\Text(139,15)[c]{{\Large $10^{\, 2}$}}
\Text(211,15)[c]{{\Large $10^{\, 3}$}}
\Text(282,15)[c]{{\Large $10^{\, 4}$}}
\Text(353,15)[c]{{\Large $10^{\, 5}$}}
\Text(31,46)[l]{{\Large $0.94$}}
\Text(31,103)[l]{{\Large $0.96$}}
\Text(31,160)[l]{{\Large $0.98$}}
\Text(31,216)[l]{{\Large $1.00$}}
\Text(31,273)[l]{{\Large $1.02$}}
\end{picture}
}
\caption{The ratio of the structure functions $ F_{2}^{\, p} $
with and without a light gluino as extrapolated from our fits
into the large-$x$, large-$Q^{2}$ region accessible at HERA.}
\end{figure}

\pagebreak

\begin{figure}[8]
\centerline{
\begin{picture}(400,420)(0,-10)
\put(0,10){\strut\epsffile{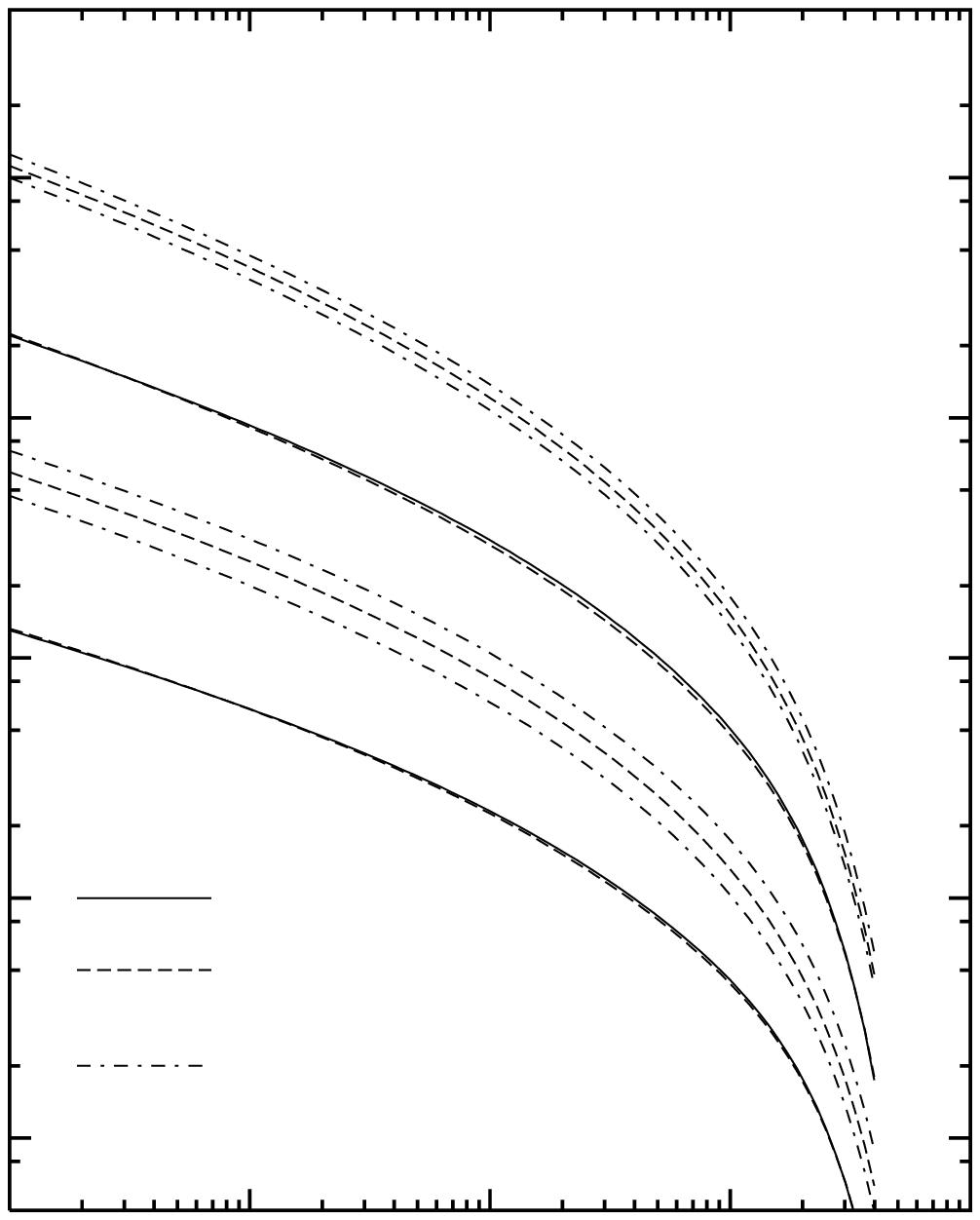}}
\Text(245,320)[l]{{\LARGE $xs, x\tilde{g}$}}
\Text(135,124)[l]{{\large no $ \tilde{g} $}}
\Text(135,102)[l]{{\large $ m_{\tilde{g}} = 0.7 \mbox{ GeV}$}}
\Text(135,74)[l]{{\large $ m_{\tilde{g}} = \!
 \begin{array}{c} {\scriptstyle 0.4} \\[-0.2cm] {\scriptstyle 1.0}
 \end{array} \! \mbox{ GeV} $}}
\Text(170,0)[l]{{\Large $x$}}
\Text(230,273)[l]{{\large $ \tilde{g} $}}
\Text(230,225)[l]{{\large $s$}}
\Text(230,195)[l]{{\large $ \tilde{g} $}}
\Text(230,146)[l]{{\large $s$}}
\Text(110,338)[l]{{$ 100 = Q^{2} (\mbox{GeV}^{2}) $}}
\Text(110,285)[l]{{$ 100 $}}
\Text(110,251)[l]{{$ 10 $}}
\Text(110,200)[l]{{$ 10 $}}
\Text(177,228)[c]{{$ (\times 0.1) $}}
\Text(177,180)[c]{{$ (\times 0.1) $}}
\Text(71,15)[c]{{\Large $10^{-4}$}}
\Text(142,15)[c]{{\Large $10^{-3}$}}
\Text(213,15)[c]{{\Large $10^{-2}$}}
\Text(285,15)[c]{{\Large $10^{-1}$}}
\Text(352,15)[c]{{\Large $1$}}
\Text(31,54)[l]{{\Large $10^{-3}$}}
\Text(31,124)[l]{{\Large $10^{-2}$}}
\Text(31,196)[l]{{\Large $10^{-1}$}}
\Text(31,266)[l]{{\Large $1$}}
\Text(31,337)[l]{{\Large $10$}}
\end{picture}
}
\caption{The radiatively generated NLO strange quark and gluino
distributions of the proton. The gluino effects on
$xs$ are shown for a fixed gluino mass, while $ x\tilde{g} $ is
given for varying gluino masses.}
\end{figure}

\end{document}